\documentclass[useAMS,usenatbib]{mn2e}
\usepackage{epsfig}

\title[The relation between column densities of interstellar OH and CH molecules]
{The relation between column densities of interstellar OH and CH molecules}
\author[T. Weselak, G.A. Galazutdinov, Y. Beletsky, and J. Kre{\l}owski]
{T. Weselak$^1$, G.A. Galazutdinov$^2$, Y. Beletsky$^3$, and J. Kre{\l}owski, J.$^4$ \\
$^1$Institute of Physics, Kazimierz Wielki University,
Weyssenhoffa 11, 85-072 Bydgoszcz, Poland;
email: towes@gazeta.pl\\
$^2$Department of Physics and Astronomy,
        19-209C, Seoul National University, Gwanak-gu, Seoul 151-747
    Republic of Korea\\
     email: runizag@gmail.com\\
$^3$ European Southern Observatory (ESO), Alonso de Cordova 3107, Santiago, Chile\\
   email:ybialets@eso.org\\
$^4$Center for Astronomy, Nicolaus Copernicus University,
Gagarina 11, Pl-87-100 Toru{\'n}, Poland; 
email: jacek@astri.uni.torun.pl  }

\begin{document}

\date{Accepted; Received ; in original form 2009}

\pagerange{\pageref{firstpage}--\pageref{lastpage}} \pubyear{2009}

\maketitle

\label{firstpage}

\begin{abstract}
We present a new, close relation between column densities of OH
and CH molecules based on 16 translucent sightlines (six of them new)
and confirm  the theoretical oscillator strengths of the OH A--X 
transitions at 3078 and 3082 \AA\ (0.00105, 0.000648) and CH B--X 
transitions at 3886 and 3890 \AA, (0.00320, 0.00210), respectively. 
We also report no difference between observed and 
previously modelled abundances of the OH molecule.

\end{abstract}


\section*{Introduction}

The interstellar OH molecule was discovered using the $\Lambda$
doublet transition observed between the levels of the ground
rotational state $^{2}\Pi_{3/2}$ J = $3/2$ at 18 cm (Weinreb et
al. 1963). Its two lines resulting from electronic transitions of
the A$^{2}\Sigma^{+}$~--~X$^{2}\Pi_{i}$ band (near 3078 and
3082~\AA), are available to ground-based observatories and were
identified in ultraviolet spectra of 14 bright OB-stars (Crutcher
and Watson 1976, Chaffee and Lutz 1977, Felenbok and Roueff 1996,
Spaans et al. 1998, Boiss\'e et al. 2005, Weselak et al. 2009b).
These findings also allowed determinations of OH column densities
along translucent sightlines. Earlier observations of the A--X
band of the OH molecule have been reanalyzed by Roueff (1996)
applying new oscillator strengths to OH A--X transitions of the
previously published line strengths. Interstellar lines of CH
molecule have been discovered by the Mount Wilson observers and
identified by McKellar (1940a, b) in spectra of OB stars due to
its A--X feature centered near 4300~\AA\ as the strongest observed
interstellar line in the violet region easily accessible to
photographic observations. However, this line is frequently
saturated. Since the B--X system near 3886~\AA\ is also quite
frequently observed and being much weaker than the A-X one it is
almost always unsaturated. Abundances of the CH molecule were
proved to be very tightly correlated with those of H$_{2}$
molecule (e.g. Mattila 1986, Weselak et al. 2004). The oscillator
strengths of CH transitions were extensively analyzed by Lien
(1984). Models of diffuse and translucent clouds have been 
extensively discussed by van Dishoeck and Black (1986).

Here we extend the above~mentioned sample (Weselak et. al 2009b) of
the UV OH absorption bands adding ground--based observations, done
using the high--resolution UVES spectrograph toward 6 new targets
(HD's: 110432, 152236, 152249, 154445, 161056, and 170740).
The aim of this work is to investigate relations between column
densities of the OH and CH molecules and test oscillator strengths
of OH A--X and CH B--X transitions based on unsaturated lines at
3078, 3082~\AA\ and 3886  and 3890~\AA, respectively. To test
oscillator strengths of the CH B--X transition we also compare results
from different instruments.

\section*{The observational data}

Most of our observational material, presented in Table 1, was
obtained using the UVES spectrograph at ESO Paranal in Chile (u
and v) with the resolution R = 80,000; one object (HD
110432) was observed in the latest observing run (v).
In the latest observing run we also observed HD's 149757 and 151932
not presented in Table 1.
In cases of objects where we found no UVES spectra we took the OH
band intensities from the literature and the intensities of CH
B--X molecular features from our own observations done using four
other spectrographs. Spectra of 4 objects were acquired using the
fiber-fed echelle spectrograph (R = 30,000, 45,000, and 90,000)
installed at the 1.8-m telescope of the Bohyunsan Optical
Astronomy Observatory (BOAO) in South Korea (b). Spectra of 2
objects was obtained using HARPS spectrometer (h), fed with the
3.6-m ESO telescope at La Silla (R = 115,000). Three objects were
observed using FEROS spectrograph (f) (R = 48,000) fed by the 2.2m
ESO telescope at La Silla and two with MAESTRO fed by the 2-m
telescope of the Observatory at Peak Terskol (t) (R = 80,000). The
spectrum of HD 34078 was also obtained with GECKO spectrograph
(g), fed with the 3.6-m Canada--France--Hawaii Telescope (CFHT)
(the spectral resolution is 120.000). Each of the above mentioned
instruments is described in the publications of Weselak et al.
(2008a, 2009b).

All the spectra were reduced using the standard packages MIDAS
and IRAF, as well as our own DECH code (Galazutdinov 1992), which
provides all the standard procedures of image and spectra
processing.  In each case the continuum placement and equivalent
width measurement, based on a
Gaussian fit, was performed as in the publication of Weselak et
al. (2009a). Due to the very low density of
interstellar clouds the only efficient line broadening mechanism
can be the Doppler one.

\section*{Results and discussion}

To obtain column densities we used the relation of Herbig (1968)
which gives proper column densities when the observed lines are
unsaturated:

\begin{equation}
N = 1.13\times10^{20} W_{\lambda}/(\lambda^{2} f),
\end{equation}

\noindent 
where W$_{\lambda}$ and $\lambda$ are in \AA\ and column
density in cm$^{-2}$. To obtain column density we adopted f-values
based on the publications of Roueff (1996) in  the case
of the A--X transition of the OH molecule and Lien (1984) in the
case of the B--X transition of the CH molecule, respectively;
(see also Weselak et al. 2009b). 
The resulting column densities of each molecule
toward the target stars are given in Table 1. In the case of OH
and CH molecules the resulting column density was obtained as a
sum of column densities originating in each line of OH and CH,
respectively. In the case of CH B--X band, which was observed by
several instruments, the resultant column densities of the CH
molecule were averaged$^{1}${\footnotetext[1]{
Column densities (in 10$^{12}$ cm$^{-2}$) of the CH molecule
in the case of multiple results in the case of HD 23180 (t): 
20.84$\pm$0.36 (based on W(3886)=4.20$\pm$0.30
and W(3890)=3.10$\pm$0.20) was averaged with that published by 
Weselak et al. (2009b)(ie. W2009b) to obtain the final column density 
equal to 21.54$\pm$1.70;
HD 24398 (t): 20.48$\pm$0.71 (based on W(3886)=4.50$\pm$0.50
and W(3890)=2.80$\pm$0.50) was averaged with that published by 
 W2009b to obtain 21.39$\pm$2.06;
HD 34078 (b): 84.27$\pm$1.92 (based on W(3886)=18.10$\pm$1.50
and W(3890)=11.80$\pm$1.20) was averaged with that published by 
 W2009b to obtain 81.77$\pm$8.31;
 HD 149757 (v and f): 25.29$\pm$0.28 and 25.13$\pm$0.71
 (based on W(3886)=5.48$\pm$0.20 and 5.50$\pm$0.50
and W(3890)=3.51$\pm$0.20 and 3.45$\pm$0.50, respectively) 
was averaged with that published by 
 W2009b to obtain 25.29$\pm$1.61;
 HD 151932 (v): 27.70$\pm$1.02 (based on W(3886)=5.75$\pm$0.41
and W(3890)=4.01$\pm$0.32);
 HD 163800 (u): 33.81$\pm$0.50 (based on W(3886)=7.10$\pm$0.30
and W(3890)=4.84$\pm$0.40) was averaged with that published by 
 W2009b to obtain 34.08$\pm$1.23}}.
 We also found new column densities of the OH molecule toward
 our target stars$^{2}$.{\footnotetext[2]{
 Column densities (in 10$^{12}$ cm$^{-2}$) of the OH molecule
 in the case of HD 147889 (u): 268.68$\pm$30.74 (based on W(3078)=13.80$\pm$1.74
and W(3082)=7.89$\pm$1.68);
HD 148688 (u): 24.73$\pm$9.88 (based on W(3078)=1.10$\pm$0.45
and W(3082)=0.81$\pm$0.54);
HD 149757 (v): 43.62$\pm$6.04 (based on W(3078)=2.10$\pm$0.67
and W(3082)=1.35$\pm$0.32);
HD 151932 (v): 71.49$\pm$10.25 (based on W(3078)=3.61$\pm$0.42
and W(3082)=2.13$\pm$0.26);
HD 152270 (u): 45.01$\pm$12.81 (based on W(3078)=1.95$\pm$0.64
and W(3082)=1.50$\pm$0.70);
HD 163800 (u): 78.43$\pm$14.27 (based on W(3078)=3.69$\pm$0.82
and W(3082)=2.31$\pm$0.78);
HD 169454 (u): 101.01$\pm$13.17 (based on W(3078)=5.14$\pm$0.85
and W(3082)=2.99$\pm$0.72).
 }
As already mentioned we used the
unsaturated bands of the selected molecules.

We examined intensity ratio of the OH features at 3078 and 3082 \AA\
based on 16 sightlines. Since their
wavelengths are very close, the strength ratio should be very
similar to that of their oscillator strengths. The latter,
according to Felenbok and Roueff (1996), are equal to 1.05(-3) and
6.48(-4) respectively. Our new result (1.62$\pm$0.03 with correlation
coefficient 0.99) demonstrates that the
equivalent width ratio matches exactly that of the oscillator
strengths confirming thus that our measurements are precise. 
We also tested the relation between equivalent widths of
unsaturated 3886 and 3890~\AA\ lines of CH as measured in spectra
from 6 different instruments. Our fit to data-points (1.43$\pm$0.05 with
correlation coefficient 0.99) is
consistent with the theoretical intensity ratio based on the
oscillator strengths published by Lien (1984). This suggests that
the intensity ratio of unsaturated 3886 and 3890~\AA\ lines of CH
is close to the expected theoretical value (1.50) suggested by
Lien (1984).

We also found new relation between column densities of the OH
and CH molecules based on our observations. New fit to data-points based
on our 16 results (N(OH)=2.86$\pm$0.12 N(CH) -- 11.41$\pm$4.73 (in 10$^{12}$ cm$^{-2}$))
with correlation coefficient equal to 0.99)
is close to previous relation presented by Weselak et al. (2009b), i.e.
(N(OH)=2.62$\pm$0.17 N(CH) -- 6.75$\pm$7.85 (in 10$^{12}$ cm$^{-2}$))
based on 10 data-points.

In Table 2 we compare column density ratios presented by Liszt and Lucas (2002)
with average ratios calculated on the basis results presented in this work.
It is well seen, that OH/CH, OH/H$_{2}$ and CH/H$_{2}$ ratios are close
to the published previously. Our results on NH/H$_{2}$ and OH/NH based on 
column densities of the NH molecule presented in the publication of Weselak et al. (2009c)
and data on H$_{2}$ taken from the literature.
In the case of relations with column densities
of NH molecule our results differ from those presented by Liszt and Lucas (2002);
the column densities of the NH molecule are not correlated with other
species containing hydrogen (Weselak et al. 2009c).

In Table 3 we also compare observational results on column densities 
of neutral diatomic molecules containing hydrogen (ie. OH, CH, NH) 
with results based on calculations of van Dishoeck and Black (1986) 
toward four stars: HD 23180, 24398, 148184 and 149757.
Small differences between calculated and observed column densities 
exist in the case of the OH molecule (HD 23180 model F, 24398 model F,
149757 model A). These models were selected on the basis
of small differences between calculated and observed column densities
of the OH and CH molecules very well correlated (Weselak et al. 2009b).
No data on column densities of the OH and NH
molecules are reported toward HD 148184. In this case we compared 
column density of CH molecule taken from the publication
of Weselak et al. (2008b) with the average calculated from models A -- G
of van Dishoeck and Black (1986).
In Table 3 we present differences between calculated and observed values 
of the column density of NH molecule. Those differences
were previously extensively discussed by Meyer and Roth (1991).
The abundance of NH toward HD 24398 was calculated by Wagenblast et al. (1993).
In this model which includes the formation of NH and OH on grains
(with (n$_{H}$)$_{0}$ = 100 cm$^{-3}$, T$_{0}$ = 30 K and I$_{UV}$ = 0.08)
abundances of OH, CH and NH molecules are 42, 21, and 0.82 $\times$
10$^{12}$ cm$^{-2}$, compared to observational results (from
Table 3) equal to 40.5$\pm$4, 21.4$\pm$2.1 and 0.90 $\times$
10$^{12}$ cm$^{-2}$, respectively.
The comparison of calculated and observed column densities 
strongly supports the thesis, that NH and OH molecules are
formed in the way of dust chemistry.

Table 4 presents physical conditions toward HD 23180, 24398 and 149757
based on comparison of our observational results with those
calculated by van Dishoeck and Black (1986) -- see also Table 3
in this work. 
We stress, that only in the case of OH molecule no difference
exist between calculated and observed column densities.

\section*{Conclusions}

The above considerations led us to infer the following
conclusions:
\begin{enumerate}
\item{
We confirm
oscillator strengths of the OH A--X transitions at
3078 and 3082 \AA\ (0.00105, 0.000648) and CH B--X transitions at
3886 and 3890 \AA, (0.00320, 0.00210), respectively. \\
}

\item{
We report no difference between observed and calculated 
column densities of OH molecule toward HD 23180, 24398 and 149757
(based on selected models of van Dishoeck and Black 1986).}

\end{enumerate}

\section*{acknowledgements}
JK and TW acknowledge financial support by the Polish State during
2007 - 2010 (grant N203 012 32/1550).
We are greateful
to an anonymous referee for valuable suggestions that allowed
us to improve the manuscript.

\newpage

\begin{table*}
\tiny{ \caption{Measured equivalent widths of interstellar features and calculated
column densities of the OH and CH molecules toward six new targets in the case
of which the OH transitions at 3078 and 3082 \AA\ were detected. 
Instruments are designated as follows: f -- FEROS, u, v -- UVES.
In the columns (9), (10), (11) we present
column densities of the OH, CH molecule based on B--X  transition, and total
column density of the CH molecule averaged in the case of multiple results, respectively
(see also text).
In the column (12) we present data on column densities of H$_{2}$ with references:
a~--~Rachford et al. (2002), b~--~Rachford et al. (2009). Data on H$_{2}$ toward
other targets is presented in the publication of Weselak et al. (2009b).
}

\begin{tabular}{rlrrrrrrrrrrrrr}
\hline\hline
HD  &   Obs &   Spec/L  &   EBV &      OH           &   OH          &   CH          &   CH          &   N(OH)           &      N(CH--BX)        &   N(CH) & N(H$_{2}$)\\
    &       &       &       &   W(3078)         &   W(3082)         &   W(3886)         &   W(3890)         &               &               &          &  \\
        &               &               &               &      [m\AA]                   &   [m\AA]          &   [m\AA]          &   [m\AA]          &    [10$^{12}$ cm$^{-2}$]  &  [10$^{12}$ cm$^{-2}$]    &  [10$^{12}$ cm$^{-2}$] & [10$^{20}$ cm$^{-2}$]\\
       (1)  & (2) & (3) & (4) & (5) & (6) & (7) & (8) & (9) & (10) & (11) & (12) \\
\hline
110432  &   f   &   B2pe    	&   0.48    &   1.81    $\pm$   0.41    &   1.28    $\pm$   0.34    &   4.10    $\pm$   0.20    &   2.40    $\pm$   0.20    &   39.72   $\pm$   6.22    &   18.12   $\pm$   0.28    & 17.91 $\pm$   0.61   & 4.37$^{a}$ \\
        &   v   &       	&       &               &               &   3.88    $\pm$   0.20    &   2.43    $\pm$   0.30    &               &   17.71   $\pm$   0.36    &           \\
152236  &   u   &   B1Iape  	&   0.66    &   3.50    $\pm$   0.44    &   2.41    $\pm$   0.34    &   5.34    $\pm$   0.45    &   4.15    $\pm$   0.56    &   75.62   $\pm$   4.30    &   27.24   $\pm$   0.72    & 27.05 $\pm$   3.10   & 5.37$^{b}$ \\
        &   f   &       	&       &               &               &   5.10    $\pm$   0.80    &   4.20    $\pm$   0.60    &               &   26.86   $\pm$   1.00    &           \\
152249  &   u   &   O9Ib    	&   0.48    &   2.81    $\pm$   1.00    &   1.82    $\pm$   0.80    &   3.40    $\pm$   0.50    &   2.61    $\pm$   0.40    &   58.61   $\pm$   14.64   &   17.23   $\pm$   1.39    & 17.23 $\pm$   1.39   & \\
154445  &   u   &   B1V 	&   0.35    &   2.01    $\pm$   0.31    &   1.52    $\pm$   0.33    &   4.06    $\pm$   0.20    &   2.84    $\pm$   0.20    &   45.92   $\pm$   10.98   &   19.59   $\pm$   0.39    & 19.59 $\pm$   0.39   & \\
161056  &   u   &   B1V 	&   0.60    &   8.38    $\pm$   0.77    &   5.30    $\pm$   0.67    &   12.70   $\pm$   0.50    &   8.80    $\pm$   0.40    &   172.46  $\pm$   12.31   &   60.98   $\pm$   1.39    & 60.98 $\pm$   1.39   & \\
170740  &   u   &   B2V 	&   0.45    &   1.97    $\pm$   0.43    &   1.73    $\pm$   0.31    &   5.10    $\pm$   0.29    &   2.84    $\pm$   0.20    &   47.29   $\pm$   9.15    &   22.02   $\pm$   0.57    & 22.02 $\pm$   0.57   & 7.24$^{a}$ \\
\hline
\end{tabular}
}
\end{table*}

\begin{table*}
\caption{
Column density ratios of diatomic molecules presented by Liszt and Lucas (2002)
compared with averaged values obtained  in this work.
Column densities of NH were taken from Weselak et al. (2009c); those of H$_{2}$ from Weselak et al. (2009b).
Our new results are based on: (1), 16; (2), 5; (3), 4 objects.
}
\begin{tabular}{lcccc}
\hline\hline
Molecular cloumn density ratio & Liszt and Lucas (2002)	& This work & \\
	\hline
OH/CH	&	3.0$\pm$0.9			&	2.52$\pm$0.35			& (1)\\
OH/H$_{2}$	&1.0$\pm$0.2$\times$10$^{-7}$	&	1.05$\pm$0.14$\times$10$^{-7}$	& (2)\\
CH/H$_{2}$	&4.3$\pm$1.9$\times$10$^{-8}$	&	4.32$\pm$0.52$\times$10$^{-8}$	& (2)\\
\hline
NH/H$_{2}$	&1.9$\pm$0.1$\times$10$^{-9}$	&	4.30$\pm$1.12$\times$10$^{-9}$	& (3)\\
OH/NH	&	50.2$\pm$5.1			&	27.78$\pm$7.47			& (3) \\
\hline
\end{tabular}
\end{table*}

\begin{table*}
\caption{
Comparison of our results on column densities (in 10$^{12}$ cm$^{-2}$) of neutral diatomic molecules containing hydrogen (Obs.)
with those calculated by van Dishoeck and Black 1986 (letter denotes model; in the case of HD 148184 averages calculated from various models).
We designate observational data from: (a), this work; (b), Roueff (1996), (c), Felenbok and Roueff (1996); (d), Meyer and Roth (1991),
(e), Weselak et al. (2008b), (f), Weselak et al. (2009c).
}
\begin{tabular}{lcccccccc}
\hline\hline
	&	HD 23180&			&	HD 24398&		&	HD 148184&		&	HD 149757	&		\\
	&		&			&				&				& 			& \\
	&	Obs.	&		F	&	Obs.	&	F	&	Obs.	&	A--G	&	Obs.	&	A	\\
	\hline
OH	&	78$\pm$26$^{b}$&	75	&	40.5$\pm$4$^{c}$&48	&	--&		40	&	43.6$\pm$6.0$^{a}$&	43	\\
CH	&	21.5$\pm$1.7$^{a}$&	12	&	21.4$\pm$2.1$^{a}$&17	&35.3$\pm$0.2$^{e}$&	12	&	25.3$\pm$1.6$^{a}$&	20	\\
NH	&	--	&		0.77	&	0.90$^{d}$&	0.33	&	--	&	0.37	&	0.76$\pm$0.25$^{f}$&	0.40	\\
\hline
\end{tabular}
\end{table*}

\begin{table*}
\caption{
Physical conditions in diffuse clouds toward HD's 23180, 24398 and 149757 based on comparison
of observational (this work) and calculated (van Dishoeck and Black 1986) column densities of the OH and CH molecules
(selected models in Table 3). 
In each case we present the number of hydrogen atoms (in cm$^{-3}$), the temperature (in K), and the
radiation field in the cloud core, respectively.
}
\begin{tabular}{lcccc}
\hline\hline
Parameter	&	HD 23180	&	HD 24398	&	HD 149757	\\
Model           &         F             &         F             &       A  \\
	\hline
(n$_{H}$)$_{0}$ [cm$^{-3}$]	&	250	&	325	&	600	\\
T$_{0}$ [K]			&	20	&	30	&	30	\\
I$_{UV}$			&	4.0	&	3.5	&	4.0	\\
\hline
\end{tabular}
\end{table*}

\end{document}